\documentclass[twocolumn,american,prl]{revtex4}
\pdfoutput=1
\usepackage{lmodern}
\usepackage[T1]{fontenc}
\usepackage[latin9]{inputenc}
\usepackage{babel}
\usepackage{float}
\usepackage{bm}
\usepackage{amsmath}
\usepackage{amssymb}
\usepackage{graphicx}
\usepackage[unicode=true,pdfusetitle,
 bookmarks=true,bookmarksnumbered=false,bookmarksopen=false,
 breaklinks=false,pdfborder={0 0 0},pdfborderstyle={},backref=false,colorlinks=false]
 {hyperref}
\hypersetup{
 colorlinks=true, citecolor=blue, linkcolor=blue, urlcolor=blue}

\makeatletter
\@ifundefined{textcolor}{}
{%
 \definecolor{BLACK}{gray}{0}
 \definecolor{WHITE}{gray}{1}
 \definecolor{RED}{rgb}{1,0,0}
 \definecolor{GREEN}{rgb}{0,1,0}
 \definecolor{BLUE}{rgb}{0,0,1}
 \definecolor{CYAN}{cmyk}{1,0,0,0}
 \definecolor{MAGENTA}{cmyk}{0,1,0,0}
 \definecolor{YELLOW}{cmyk}{0,0,1,0}
}

\makeatother

\begin{document}

\title{Valley polarized magnetic state in hole-doped mono layers of transition
metal dichalcogenides}

\author{João E. H. Braz,$^{1}$ B. Amorim,$^{1}$ Eduardo V. Castro$^{1,2}$}

\affiliation{$^{1}$CeFEMA, Instituto Superior T\'{e}cnico, Universidade de Lisboa,
Av. Rovisco Pais, 1049-001 Lisboa, Portugal}

\affiliation{$^{2}$Beijing Computational Science Research Center, Beijing 100084,
China}
\begin{abstract}
We compute the valley/magnetic phase diagram of mono layers of transition
metal dichalcogenides in the hole doped region where spin-orbit effects
are particularly relevant. Taking into account the moderate to high
local electron-electron interactions due to the presence of transition
metal atoms, we show that the system is unstable to an itinerant ferromagnetic
phase where all charge carriers are spin and valley polarized. This
phase shows an anomalous charge Hall and anomalous spin-Hall response,
and may thus be detected experimentally. 
\end{abstract}
\maketitle


\emph{Introduction.\textemdash }The manipulation of quantum degrees
of freedom such as the electron charge and spin is an essential ingredient
in the development of better devices and novel quantum technologies.
Yet charge and spin do not exhaust all quantum degrees of freedom
associated with electrons in solids. In certain semiconductors, multiple
degenerate valence band maxima or conduction band minima \textendash{}
the so called \emph{valleys} \textendash{} occur. Carriers are then
characterized not only by charge and spin, but also by the \emph{valley}
degree of freedom indicating the region in momentum space where they
are confined. Valleytronics aims to use this quantum degree of freedom
in novel technological devices in much the same way the electron spin
is used in spintronics. Individual valley manipulation is, however,
a necessary requirement. Well known semiconductors such as Si and
Ge display valleys \cite{Ashcroft}, but it is difficult to have an
external coupling to a single valley in these cases, which severely
limits their manipulation. 

The isolation of real 2D materials with hexagonal lattice like graphene
have put valley physics on the spotlight again. The manipulation of
the two valleys of graphene is, however, not easy to achieve \cite{NGPrmp}.
Such manipulation has finally been clearly demonstrated for the new
class of 2D materials known as \emph{semiconducting transition metal
dichalcogenides} (TMDs) \cite{XXH14}, formula MX$_{2}$, where M
is a transition metal (ex. Mo, W) and X is a chalcogen (ex. S, Se)
\cite{wang2012electronics,yazievKis15}. The demonstration that electrons
of either valley can be excited across the gap by conveniently choosing
either left or right circularly polarized light paves the way to valleytronic
devices \cite{heinzNatNanot12,zeng2012valley,CWH+12}. The situation
is even more interesting because of the non-negligible spin-orbit
coupling in TMDs, which induces a sizable valence band spin-splitting
\cite{xiao2012}. Valley polarization may then be achieved by applying
a perpendicular magnetic field \cite{heinzSplit14,MHM+15,AGJ15,SSAL+15,QLN+15},
which combined with optical absorption makes coherent valley manipulation
possible \cite{YSHnatPhys2016,urbaszek2016,potemski2016,schmidt2016magnetic}.
\begin{figure}[H]
\begin{centering}
\includegraphics[width=8cm]{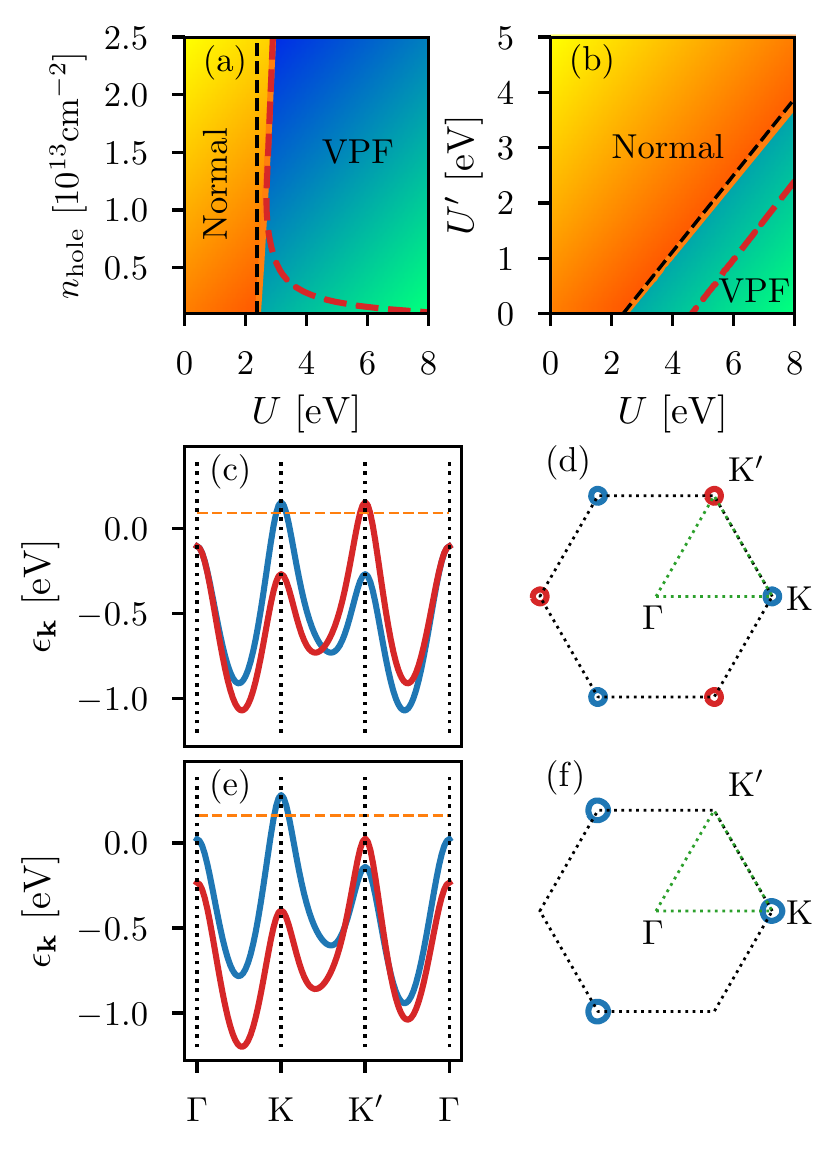}
\par\end{centering}
\caption{\label{fig:pd}(a) Mean field phase diagram in the $n_{\text{hole}}-U$
plane, indicating the normal and the valley polarized ferromagnetic
(VPF) phases, at a temperature of $T=1$ K and $U^{\prime}=0$ for
the TMD WS$_{2}$. The dashed red line indicates the transition at
$T=100$ K. The vertical dashed line represents the estimated critical
$U$ of $2.38$~eV in the limit of low $T$ and $n_{\text{hole}}$
obtained using a low energy model. (b) Phase diagram in the $U-U^{\prime}$
plane at a hole density of $n_{\text{hole}}=0.2\times10^{13}\text{ cm}^{-2}$
and $T=1$ K. The dashed red line, shows the transition line at the
temperature of $T=100$ K. The dashed black line indicates the critical
line $U=U_{c}+1.4U^{\prime}$, estimated using the low energy model.
Panels (c) and (e) show the band structure for the spin up (in red)
and spin down (in blue) valence bands in the normal and VPF phases,
respectively. The horizontal dashed line indicates the Fermi level
for a constant particle number of $n_{\text{hole}}=10^{13}\text{ cm}^{-2}$.
The Fermi surfaces for the normal and VPF phases are represented in
panels (d) and (f), respectivelly.}
\end{figure}

For systems with degenerate valleys like TMDs, an obvious question
is whether a spontaneously valley polarized phase can emerge. The
question is not only of fundamental interest, since possible device
applications in valleytronics will require valley polarized materials
in as much the same way spintronic devices require materials with
long range magnetic order. Spontaneous valley polarization was predicted
more than thirty years ago for Si inversion layers \cite{BSV79},
and experimentally confirmed soon after \cite{cole1981evidence}.
This phase has nevertheless been elusive within the much recent field
of novel 2D materials, and only in the Landau level regime it has
some relevance \cite{FKSY12,WAF+10,AYoungBil2016,ghaemi2012fractional,abanin2012interaction}.
Only recently have spontaneous valley polarization been used to explain
magnetoluminescence results in electron doped WS$_{2}$ \cite{scrace2015magnetoluminescence}. 

In this work we predict that hole doped TMDs, in particular those
with large spin-splitting of the valence band like WS$_{2}$, display
a valley polarized ferromagnetic (VPF) phase which should be robust
in a wider range of parameters than its electron doped counterpart
\cite{scrace2015magnetoluminescence}. A typical phase diagram is
shown in Figs.~\ref{fig:pd}(a) and (b) in the plane of intra-orbital
$U$ Coulomb interaction and hole density and, $U$ and inter-orbital
$U^{\prime}$ interaction, respectively. Even though the Coulomb repulsion
parameters are largely unknown for TMDs, both the $T=1$ K (full line)
and the $T=100$ K (dashed line) transition lines put the valley polarized
phase within reach according to current parameter estimates \cite{RCG13}.
As shown Figs.~\ref{fig:pd}(c) and (d) in the normal phase there
are degenerate hole pockets at both inequivalent valleys $K$ and
$K^{\prime}$. Even though in each valley the carriers are spin polarized
due to spin-orbit interaction, since the two valleys are degenerate
there is no net spin polarization. On the contrary, in the valley
polarized phase only a single valley is occupied, as shown in Figs.~\ref{fig:pd}(e)
and (f). The system then realizes a valley polarized ferromagnet with
a single Fermi pocket occupied. A key ingredient for this complete
valley and spin polarization is the large spin-splitting of the valence
band due to strong spin-orbit coupling. For the conduction band, the
spin-splitting is smaller by one order of magnitude at least \cite{rossierHetero2013},
given rise to a partially valley polarized \cite{scrace2015magnetoluminescence},
and thus less stable phase.


\emph{Model and variational mean field treatment.\textemdash }We model
electrons in TMDs using a M atom 3-orbital nearest neighbor tight-binding
Hamiltonian. The free part of the Hamiltonian is given by 
\begin{equation}
H_{0}=\sum_{\left\langle i,j\right\rangle }\sum_{\gamma,\gamma',\sigma}c_{i,\gamma,\sigma}^{\dagger}E_{\gamma,\gamma'}^{\sigma}(\mathbf{r}_{ij})c_{j,\gamma',\sigma}+H_{SO}\,,\label{eq:Hamiltonian_TB}
\end{equation}
where $c_{i,\gamma,\sigma}^{\dagger}$ is a electron creation operator
on lattice site $i$, M atom orbital $\gamma=d_{z^{2}},d_{xy},\,d_{x^{2}-y^{2}}$
and spin $\sigma=\uparrow,\downarrow$. $E_{\gamma,\gamma'}^{\sigma}(\mathbf{r}_{ij})$
are hopping integrals as given in Ref.~\cite{xiao3bTB} for the nearest
neighbor model \footnote{We use the hopping integrals obtained from the GGA DFT calculation
of Ref.~\cite{xiao3bTB}}, and $H_{SO}$ is the spin-orbit coupling term. For the interacting
part of the Hamiltonian, we consider on-site interactions, including
the intra- ($U$) and inter-orbital ($U^{\prime}$) Coulomb interactions,
as well as Hund ($J$) and pair-hopping ($J^{\prime}$) terms:
\begin{align}
H_{\text{int}} & =\frac{U}{2}\sum_{\begin{subarray}{c}
i,\gamma\\
\sigma\neq\sigma^{\prime}
\end{subarray}}n_{i,\gamma,\sigma}n_{i,\gamma,\sigma^{\prime}}+\frac{U^{\prime}}{2}\sum_{\begin{subarray}{c}
i,\gamma\neq\gamma^{\prime}\\
\sigma,\sigma^{\prime}
\end{subarray}}n_{i,\gamma,\sigma}n_{i,\gamma^{\prime},\sigma^{\prime}}\nonumber \\
 & +\frac{J}{2}\sum_{\begin{subarray}{c}
i,\gamma\neq\gamma^{\prime}\\
\sigma,\sigma^{\prime}
\end{subarray}}c_{i,\gamma,\sigma}^{\dagger}c_{i,\gamma^{\prime},\sigma^{\prime}}^{\dagger}c_{i,\gamma,\sigma^{\prime}}c_{i,\gamma^{\prime},\sigma}\nonumber \\
 & +\frac{J^{\prime}}{2}\sum_{\begin{subarray}{c}
i,\gamma\neq\gamma^{\prime}\\
\sigma\neq\sigma^{\prime}
\end{subarray}}c_{i,\gamma,\sigma}^{\dagger}c_{i,\gamma,\sigma^{\prime}}^{\dagger}c_{i,\gamma^{\prime},\sigma^{\prime}}c_{i,\gamma^{\prime},\sigma}\,,
\end{align}
where we have written $n_{i,\gamma,\sigma}=c_{i,\gamma,\sigma}^{\dagger}c_{i,\gamma,\sigma}$.
From the four parameters characterizing the on-site interaction only
two are independent as by symmetry arguments one has for $d$ orbitals
$J^{\prime}=J=\left(U-U^{\prime}\right)/2$ \cite{DHM01}.

We performed a mean field analysis of the interacting Hamiltonian
$H=H_{0}+H_{\text{int}}$. We focused on homogeneous phases that are
diagonal in the spin and orbital degrees of freedom. The mean field
Hamiltonian thus reads
\begin{equation}
H_{MF}=H_{0}+\sum_{\begin{subarray}{c}
\gamma,\sigma\end{subarray}}\phi_{\gamma,\sigma}\sum_{i}n_{i,\gamma,\sigma}\,,\label{eq:hamilt}
\end{equation}
where $\phi_{\gamma,\sigma}$, the molecular fields for atomic orbital
$\gamma$ and spin $\sigma$, constitute variational parameters. Due
to the spin-valley coupling in TMDs, we have that magnetic instabilities,
which break time reversal symmetry, also lift valley degeneracy. We
therefore focus on magnetic phases and assume a minimum set of variational
parameters, with
\begin{equation}
\phi\equiv\phi_{\uparrow}\equiv\phi_{d_{z^{2}},\uparrow}=\phi_{d_{xy},\uparrow}=\phi_{d_{x^{2}-y^{2}},\uparrow}=-\phi_{\downarrow}.
\end{equation}
This ansatz corresponds to a relative shift in energy of the spin
up and down states keeping the same electronic dispersion relation
for each spin component. We analyze the possible phases at fixed hole
concentration, $n_{\text{hole}}$, by minimizing the mean field functional
\begin{equation}
\mathcal{F}\left[\phi\right]=\Omega_{MF}+\mu\left\langle N_{e}\right\rangle _{MF}+\langle H-H_{MF}\rangle_{MF},\label{eq:feVar}
\end{equation}
where $\Omega_{MF}=-k_{B}T\log\text{Tr}\left\{ e^{-\beta\left(H_{MF}-\mu N_{e}\right)}\right\} $
is the grand potential for the mean field Hamiltonian, with $N_{e}$
the total electron number operator, and $\langle...\rangle_{MF}$
the thermodynamical average with respect to $H_{MF}$. In the previous
equation, the chemical potential is determined by the condition
\begin{equation}
2-n_{\text{hole}}=\frac{1}{N}\sum_{\mathbf{k},n,\sigma}f\left(\epsilon_{\mathbf{k},n,\sigma}^{MF}-\mu\right),
\end{equation}
where $N$ is the number of lattice sites, $n_{\text{hole}}$ is the
density of holes per unit cell, $f(\epsilon)=\left(e^{\beta\epsilon}+1\right)^{-1}$
is the Fermi-Dirac function and $\epsilon_{\mathbf{k},n,\sigma}^{MF}$
are the bands of $H_{MF}$ \footnote{We used the fact that the in the neutral system there are two electrons
per unit cell: one with spin up and other with spin down.}.

\begin{figure}
\begin{centering}
\includegraphics[width=8cm]{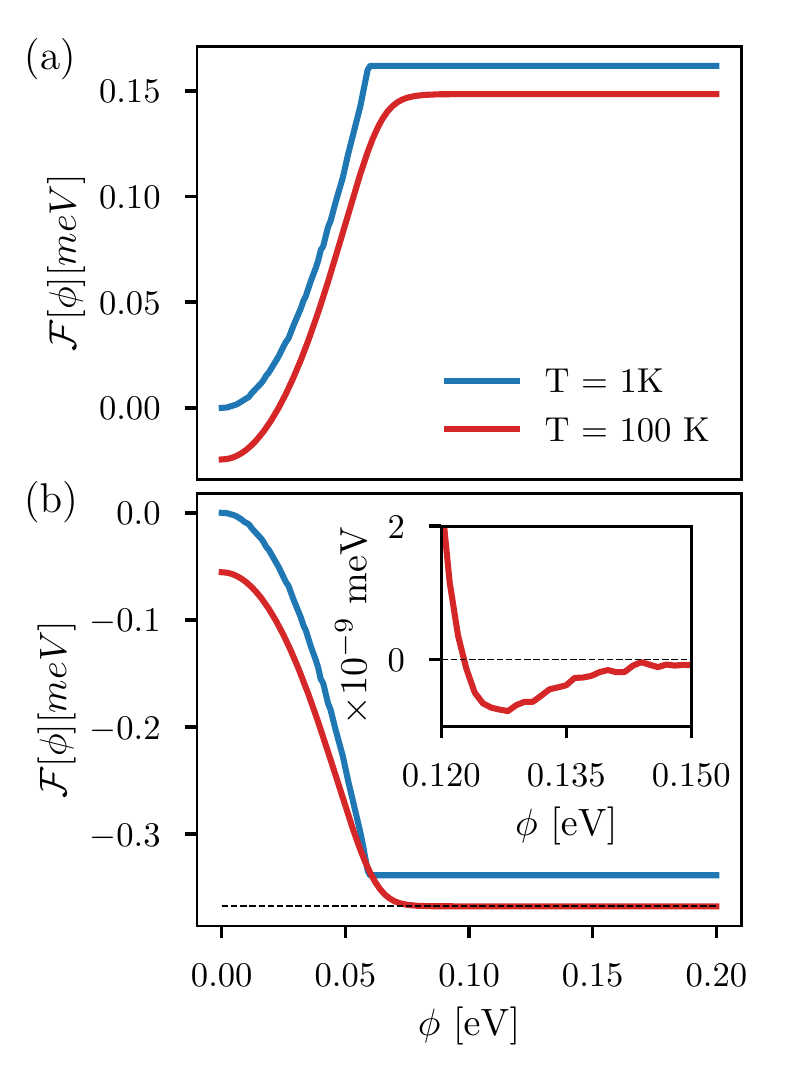}
\par\end{centering}
\caption{\label{fig:fe-eq}Variational mean field functional, $\mathcal{F}\left[\phi\right]$,
as a function of the variational parameter $\phi$ for WS$_{2}$ at
fixed $n_{\text{hole}}=10^{13}\text{ cm}^{-2}$ and two different
temperatures: $T=1$~K (blue) and $T=100$~K (red). Panel~(a) shows
the functional in the normal phase region ($U=1$ eV and $U^{\prime}=0$)
and panel (b) in the VPF phase region ($U=6$ eV and $U^{\prime}=0$).
The inset in panel~(b) zooms the $T=100$~K curve, showing the occurrence
of the minimum in $\mathcal{F}\left[\phi\right]$ indicating the stability
of the VPF phase.}
\end{figure}

\begin{figure}
\begin{centering}
\includegraphics[width=8cm]{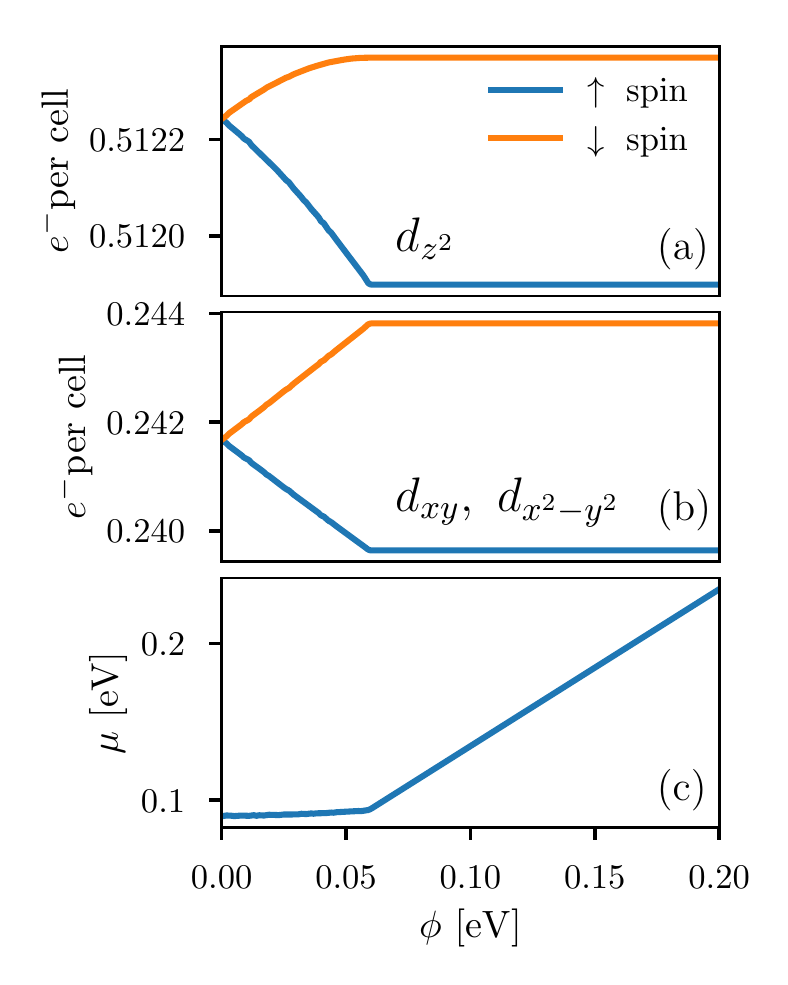}
\par\end{centering}
\caption{\label{fig:mu-eq}Spin and orbital resolved electron densities as
a function of the variational parameter $\phi$ for (a) the $d_{z^{2}}$
orbital and for (b) the degenerate $d_{xy}$ and $d_{x^{2}-y^{2}}$
orbitals. Panel (c) shows the evolution of the chemical potential,
$\mu$, as a function of the variational parameter.}
\end{figure}

In the following, we focus on the large valence band spin-splitting
TMD, WS$_{2}$, for hole dopings where the chemical potential lies
within the spin-splitting of the valence bands. In Figs.~\ref{fig:pd}(a-b),
we show the phase diagram of WS$_{2}$. As can be seen in In Fig.~\ref{fig:pd}(a),
for a given hole density $n_{\text{hole}}$ and $U^{\prime}$ value,
there is a critical value of $U$ above which the system goes into
a valley polarized ferromagnetic phase. Using a low energy model,
which correctly captures interactions between holes and the full band
as well as the multi-orbital character of the system, we estimate
$U_{c}\simeq2.38$ eV in the limit of low temperature and hole density
with $U^{\prime}=0$. For finite $U^{\prime}$, the low energy model
predicts that the system will be ferromagnetic provided $U>U_{c}+1.4U^{\prime}$,
which is in good agreement with the tight-binding results shown in
Fig.~\ref{fig:pd}(b). As represented in Fig.~\ref{fig:pd}(c-f),
in the VPF phase the system becomes fully valley and spin polarized,
with one of the spin polarized bands becoming fully occupied, while
the opposite polarized band remaining partially occupied. This is
further shown in Fig.~\ref{fig:fe-eq}, where we plot the behavior
of the mean field functional $\mathcal{F}\left[\phi\right]$ as a
function of $\phi$ for a hole density of $n_{\text{hole}}=10^{13}\,\mbox{cm}^{-2}$.
The plateau region seen in $\mathcal{F}\left[\phi\right]$ signals
the case where the system becomes fully valley and spin polarized,
with one of spin polarized bands becoming fully occupied, while the
opposite polarized band remains partially occupied. In the zero temperature
limit, once one of the spin bands becomes fully occupied $\mathcal{F}\left[\phi\right]$
no longer depends on $\phi$. At finite temperature, there will always
be some hole density in the minority band and therefore $\mathcal{F}\left[\phi\right]$
will have a weak dependence on $\phi$ as shown in the inset of Fig.~\ref{fig:fe-eq}(b),
where, within numerical precision, the minimum of $\mathcal{F}\left[\phi\right]$
indicating the stability of the VPF phase can be seen. The spin and
orbit resolved densities are plotted in Fig.~\ref{fig:mu-eq}(a-b)
for the case of $U=6\,\mbox{eV}$ and $U^{\prime}=0\,\mbox{eV}$,
at $n_{\text{hole}}=10^{13}\,\text{cm}^{-2}$ and $T=1$ K. As can
be seen in Fig.~\ref{fig:mu-eq}(a) for the $d_{z^{2}}$ orbital,
and in Fig.~\ref{fig:mu-eq}(b) for the degenerate orbitals $d_{xy}$
and $d_{x^{2}-y^{2}}$, the system becomes a spin polarized metal.
The evolution of the chemical potential is shown in panel (c). Similar
results are obtained for other TMDs of the semiconducting family.

\emph{Anomalous Hall and other responses.\textemdash{}} Although TMDs
possess a bands with a locally non-vanishing Berry curvature, intrinsic
time-reversal symmetry (TRS) voids them of an anomalous Hall (AH)
response other than the valley Hall effect~\cite{xiao2012}. The
spontaneous breaking of TRS in an itinerant magnetic phase provides
a richer response, in virtue of the simultaneous polarization in the
spin and valley degrees of freedom. 

The AH conductivity can be expressed in terms of the Berry connection
as \cite{XCN10}
\begin{equation}
\sigma^{AH}=-\frac{e^{2}}{\hbar}\frac{1}{A_{c}N}\sum_{\mathbf{k},n,\sigma}f\left(\epsilon_{\mathbf{k},n,\sigma}^{MF}-\mu\right)\Omega_{\mathbf{k},n,\sigma},
\end{equation}
where $\Omega_{\mathbf{k},n,\sigma}$ is the Berry curvature and $\epsilon_{\mathbf{k},n,\sigma}^{MF}=\epsilon_{\mathbf{k},n,\sigma}^{0}+\sigma\phi$
is the mean-field dispersion relation, with $\epsilon_{\mathbf{k},n,\sigma}^{0}$
the eigenenergies of Eq.~\eqref{eq:Hamiltonian_TB}, and we use $A_{c}$
the unit cell area. For hole doping, we can write the AH conductivity
as a contribution from the full bands $\bar{\sigma}^{AH}$ and a contribution
from the holes $\sigma_{\text{h}}^{AH}$, i.e. $\sigma^{AH}=\bar{\sigma}^{AH}+\sigma_{\text{h}}^{AH}$,
where the full band contribution is null. For weak doping, the main
contribution to $\sigma_{\text{h}}^{AH}$ comes from the valence band
pockets at the $K$ and $K^{\prime}$ points, which have, respectively,
spin up and spin down polarizations. Therefore we can write $\sigma_{\text{h}}^{AH}\simeq\sigma_{\text{h},+,\uparrow}^{AH}+\sigma_{\text{h},-,\downarrow}^{AH}$,
where
\begin{equation}
\sigma_{\text{h},\tau,\sigma}^{AH}=\frac{e^{2}}{\hbar}\frac{1}{A_{c}N}\sum_{\mathbf{k}}\left[1-f\left(\epsilon_{\mathbf{k},\tau,v,\sigma}-\mu\right)\right]\Omega_{\mathbf{k},\tau,v,\sigma}\,,
\end{equation}
with $v$ indicating valence band. Neglecting the momentum dependence
of the Berry curvature, which is valid in the limit of small doping,
we approximate
\begin{equation}
\sigma_{\text{h},\left(+,\uparrow\right)/(-,\downarrow)}^{AH}\simeq\pm\frac{e^{2}}{\hbar}\Omega_{0}n_{\text{hole},\uparrow/\downarrow},
\end{equation}
where $n_{\text{hole},\uparrow/\downarrow}$ is the spin up/down hole
density per area and $\Omega_{0}=\Omega_{K,v,\uparrow}=-\Omega_{-K,v,\downarrow}$
is the Berry curvature at the $K/K^{\prime}$ point. The Berry curvature
at the $K$ and $K^{\prime}$ points can be computed within a low
energy, continuum model. Using $k\cdot p$ theory, we can write an
effective two-band model valid close to the $\tau K$ point
\begin{align}
\bm{H}_{\mathbf{k},\tau,\sigma} & =\left(\mathcal{E}_{0}+\sigma\tau\frac{\lambda}{2}+\alpha a^{2}k^{2}\right)\bm{I}\nonumber \\
 & +\left(\Delta-\sigma\tau\frac{\lambda}{2}+\beta a^{2}k^{2}\right)\sigma_{z}+ua\bm{\sigma}_{\tau}\cdot\mathbf{k},
\end{align}
where $\bm{\sigma}_{\tau}=\left(\tau\sigma_{x},\,\sigma_{y}\right)$
and $\mathcal{E}_{0}\simeq0.84$~eV, $\Delta\simeq0.9$~eV, $\alpha\simeq0.26$~eV,
$\beta\simeq0.38$~eV, and $u\simeq1.69$~eV, with $a\simeq3.191\text{\,Å}$
the lattice parameter and $\lambda\simeq0.211$~eV the spin-orbit
coupling. From the eigenstates of this Hamiltonian we can evaluate
the Berry curvature as \cite{xiao2012}, 
\begin{equation}
\Omega_{\mathbf{k},\tau,v,\sigma}=a^{2}\frac{\tau}{2}\frac{\left(ua\right)^{2}\left(\Delta-\sigma\tau\frac{\lambda}{2}-\beta a^{2}k^{2}\right)}{\left[\left(\Delta-\sigma\tau\frac{\lambda}{2}+\beta a^{2}k^{2}\right)^{2}+\left(ua\right)^{2}k^{2}\right]^{3/2}},
\end{equation}
from which we obtain $\Omega_{0}=\Omega_{\mathbf{0},+,v,\uparrow}=\Omega_{\mathbf{0},-,v,\downarrow}=a^{2}\left(ua\right)^{2}/\left[2\left(\Delta-\frac{\lambda}{2}\right)^{2}\right]\simeq20\text{ Å}^{2}$.
Therefore the AH response if proportional to the magnetization of
the system $\sigma^{AH}\simeq\frac{e^{2}}{\hbar}\Omega_{0}\left(n_{\text{hole},\uparrow}-n_{\text{hole},\downarrow}\right)$.
Besides the charge response, we can also consider spin and spin-valley
responses, which can be obtained as $\sigma_{q}^{AH}=-\sum_{\tau,\sigma}c_{q}^{\tau,\sigma}\sigma_{\text{h},\tau,\sigma}^{AH}/e$,
with $c_{s}^{\tau,\sigma}=\sigma$ for the spin Hall and $c_{sv}^{\tau,\sigma}=\sigma\tau$
for the spin-valley Hall responses. In the VPF phase, only one of
the valleys is populated with holes (for concreteness we assume that
is the $K$ point) implying $\sigma^{AH}\simeq\frac{e^{2}}{\hbar}\Omega_{0}n_{\text{hole},\uparrow}$
and $\sigma^{AH}=e\sigma_{sv}^{AH}=-e\sigma_{s}^{AH}$. We conclude
that this phase has a transversal response that is polarized in both
the spin and valley degrees of freedom.

Further responses can be qualitatively inferred from the structure
of the magnetic bands depicted in Fig.~\ref{fig:pd}(e). Besides
the transversal component computed above, the system will respond
with a longitudinal component, in virtue of its metallic state, that
is just as well spin- and valley-polarized. Moreover, optical transitions
in this phase will also be polarized, in virtue of the inequivalent
valleys having differing optical gaps. Note that the physics of the
latter response differs from that of earlier reports such as in Refs.~\cite{heinzNatNanot12,zeng2012valley,CWH+12},
since the magnetic ground-state spontaneously breaks TRS, whereas
previously this symmetry has been explicitly broken using circularly
polarized photons.


\emph{Conclusions.\textemdash }Based in the present mean field calculations,
we have shown that TMDs are unstable to a spin-valley polarized metal.
For the phase to be observed it is required that $U>U_{c}\simeq2.38\,\mbox{eV}$
and $U^{\prime}<0.7\left(U-U_{c}\right)$, which are realistic conditions
given the transition metal atoms involved. Experimentally, this phase
could be detected by the measurements of the anomalous Hall and/or
longitudinal response, both of which are spin and valley polarized.
Interestingly, the spin and valley polarization is opposite for the
two responses. Also, the presence of a valley polarized, magnetic
field tunable, positively charged exciton (X$^{+}$ trion) in the
PL spectrum of hole doped TMDs would be a clear indication by optical
means of this phase \cite{scrace2015magnetoluminescence}. Even though
defects make TMDs naturally electron doped \cite{zhu2014exciton},
holes can be induced by electric field effect and the observation
of X$^{+}$ excitations is possible \cite{ross2013electrical}.\textbf{
}The present results show that a valley polarized phase can be achieved
in TMDs without the need of an exchange coupling to a permanent magnet
\cite{QLN+15,kimBaisSVP2016}. They also agree with a recent non self-consistent
approach \cite{rostamiMag15} and with DFT calculations for a monolayer
of 2H-VSe$_{2}$ \cite{duanNCommSVP2016}.
\begin{acknowledgments}
E.C. acknowledges the financial support of FCT-Portugal through grant
No. EXPL/FIS-NAN/1728/2013. Partial support from FCT-Portugal through
Grant No. UID/CTM/04540/2013 is also acknowledged. B. Amorim received
funding from the European Union\textquoteright s Horizon 2020 research
and innovation programme under grant agreement No 706538.

\end{acknowledgments}


%

\end{document}